\documentclass[12pt]{article}
\usepackage{epsfig}
\title{\bf  Weak interaction contribution  to\\ the inclusive
hadron-hadron scattering \\cross sections at high $p_T$}
\author{ B.L.Ioffe\\
A.I.Alikhanov Institute of Theoretical and Experimental Physics\\
B.Cheremushkinskaya 25, 117218 Moscow,Russia}
\date{}
\begin{document}

\maketitle

\newcommand{\be}{\begin{equation}}
\newcommand{\ee}{\end{equation}}

\def\la{\mathrel{\mathpalette\fun <}}
\def\ga{\mathrel{\mathpalette\fun >}}
\def\fun#1#2{\lower3.6pt\vbox{\baselineskip0pt\lineskip.9pt
\ialign{$\mathsurround=0pt#1\hfil##\hfil$\crcr#2\crcr\sim\crcr}}}

\def\Journal#1#2#3#4{{#1} {#2} (#4) #3 }
\def\NCA{{\em Nuovo Cimento} A}
\def\PHYS{{\em Physica}}
\def\NPA{{\em Nucl. Phys.} A}
\def\MATH{{\em J. Math. Phys.}}
\def\PRO{{\em Prog. Theor. Phys.}}
\def\NPB{{\em Nucl. Phys.} B}
\def\PLA{{\em Phys. Lett.} A}
\def\PLB{{\em Phys. Lett.} B}
\def\PLD{{\em Phys. Lett.} D}
\def\PL{{\em Phys. Lett.}}
\def\PRL{\em Phys. Rev. Lett.}
\def\PREV{\em Phys. Rev.}
\def\PREP{\em Phys. Rep.}
\def\PRA{{\em Phys. Rev.} A}
\def\PRD{{\em Phys. Rev.} D}
\def\PRC{{\em Phys. Rev.} C}
\def\PRB{{\em Phys. Rev.} B}
\def\ZPC{{\em Z. Phys.} C}
\def\ZPA{{\em Z. Phys.} A}
\def\ANNP{\em Ann. Phys. (N.Y.)}
\def\RMP{{\em Rev. Mod. Phys.}}
\def\CHEM{{\em J. Chem. Phys.}}
\def\INT{{\em Int. J. Mod. Phys.} E}

\vspace{5mm}

\begin{abstract}

\vspace{10mm}

It is demonstrated that the strong power-like scaling violation in the transverse
momentum distribution of inclusive hadron production, observed by CDF Collaboration in
$\bar{p}p$ collisions at Tevatron is caused by contribution  of weak interaction. The
contribution of weak interaction is increasing with energy at high energies.

\end{abstract}


\maketitle

\vspace{10mm}

The CDF Collaboration  have measured the inclusive  cross sections of charged hadron
production at high transverse momentum $p_T$ at $\bar{p}p$ collisions at c.m. energy 1.96
TeV \cite{CDF}. Surprisingly the strong power-like scaling violation was observed at $p_T
> 30$ GeV: at $p_T \approx 100$ GeV the data indicate that the scaling law $Ed\sigma/d^3
p \sim 1/p_T^4$ is violated more than by one  order of magnitude. The scaling law $E
d\sigma/d^3p\sim 1/p_T^4$ for inclusive hadron production in hadron-hadron scattering was
proved basing on very general grounds -- the light-cone dominance of hard processes in
strong interaction \cite{Ioffe}. Therefore, the observation of the violation of the
scaling law resulted to strong confusion. Theoretically the observed phenomenon was
discussed in the paper by Albino et al \cite{Albino}. The authors of Ref.\cite{Albino}
addressed the scaling violation to factorization breaking at high transverse momentum
charged hadron production. Such explanation is not satisfactory: QCD has no scale
parameters besides $\Lambda_{QCD}$ and inclusive cross sections are infra-red stable in
QCD. In principal there is the dimensional parameter in the problem in view -- the energy
of the collision. But, as it is well known \cite{Ioffe},\cite{Gribov}, the energy is
related to the longitudinal size of the collision region, but not to the transverse size,
which determines the cross section. In recent paper \cite{Yoon} an attempts were done to
constract the models, describing the data, but as well as in \cite{Albino} no success was
achieved. At the same time the measurements of inclusive jet production
\cite{Aaltonen},\cite{Abazov} demonstrate good agreement with scaling law and theoretical
expectations.

In this paper it is shown, that the scaling law violation in inclusive cross sections of
charged hadron production at high $p_T$, observed by CDF Collaboration, is  described by
contribution of weak interaction. The idea is that weak interaction has the scale
parameters -- the masses of $W$ and $Z$ bosons. At high $p_T$ the contribution of weak
interaction to the inclusive cross section is strongly enhanced by the presence of $W$
and $Z$ resonances  in comparison with strong interaction contribution which falls
steeply with $p_T$. Weak interaction contribution   has a peak at $p_T =m_W/2$. Due to
these circumstances the weak interaction contribution becomes compatible with strong ones
at $p_T \ga 30 $ GeV. The weak interaction Lagrangian is the ones of the Standard Model:
$$ L = \frac{g}{\sqrt{2}} \left\{ \biggl [ W_{\mu}^+ \bar{u} \gamma_{\mu} \frac{1}{2}
(1+\gamma_5)d +W^-_{\mu} \bar{d} \gamma_{\mu}  \frac{1}{2}(1+\gamma_5) u\biggr ]~ +
\right.$$
$$ + ~Z_{\mu} \frac{1}{\mbox{cos} \theta_W} \biggl [ \bar{u} \gamma_{\mu} \frac{1}{2}
(1+\gamma_5)\biggl ( \frac{1}{2} -\frac{2}{3} \mbox{sin}^2 \theta_W\biggr ) u + \bar{u}
\gamma_{\mu}\frac{1}{2} (1-\gamma_5) \biggl ( -\frac{2}{3} \mbox{sin}^2 \theta_W\biggr )u
+$$ \be \left.+\bar{d}\gamma_{\mu} \frac{1}{2} (1+\gamma_5) \biggl ( -\frac{1}{2}
+\frac{1}{3} \mbox{sin}^2\theta_W \biggr )d +\bar{d} \gamma_{\mu} \frac{1}{2}
(1+\gamma_5) \biggl ( \frac{1}{3} \mbox{sin}^2 \theta_W\biggr )d\biggr ] + (u \to c,~d\to
s)\right\}.\label{1}\ee Here $u$ and $d$ are fields of $u$ and $d$ quarks, $\theta_W$ is
the Weinberg angle, sin$^2 \theta_W \approx$ 0.230. The coupling constant $g$ is equal
\be g^2=\frac{e^2}{\mbox{sin}^2\theta_W},~~~e^2 =\frac{1}{137}\label{2}\ee The matrix
element of weak interaction contribution to the inclusive cross section in $\bar{p}p$
collision is represented by the diagram of Fig.1

\bigskip

\begin{figure}[h]
\hspace{50mm} \epsfxsize=6.0cm \epsfbox{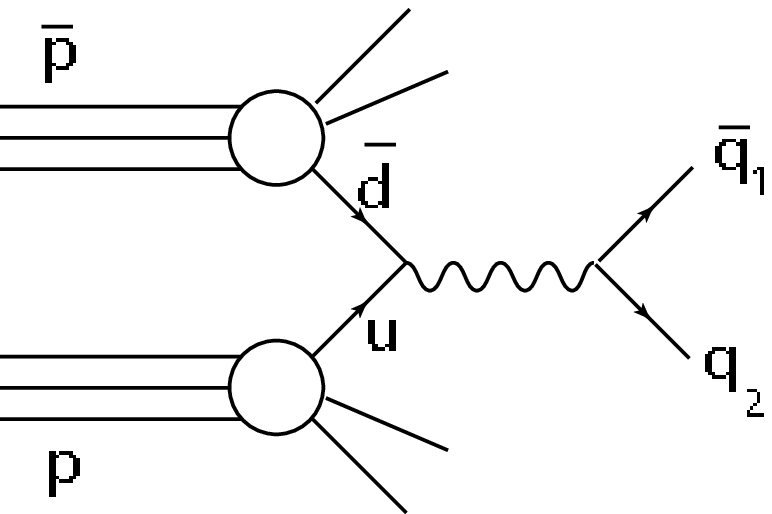}


\vspace{3mm}

{\bf Fig.1.}  The diagram, describing the quark pair production at high $p_T$ in case of
$W^+$ exchange in annihilation channel.

\end{figure}

 There are also the diagrams, where $\bar{p}$ fragments into $u,\bar{s}, c$ and
$p$ -- into $\bar{d},c,\bar{s}$, correspondingly, as well the diagrams with $W^-$ and $Z$
in annihilation channel. (The contribution of $W$ and $Z$ exchange in $t$-channel is
negligable.) In order to compare the results with CDF data let us calculate the inclusive
cross section integrated over pseudorapidity
\be \eta =\frac{1}{2} \ln \frac{E' + p'_{\parallel}}{E' - p'_{\parallel}},\label{3}\ee
where $E^{\prime}= \sqrt{p ^{\prime 2}_{\parallel} + p ^{\prime 2}_T}$ is the energy of
detected charged particle, $p'=(p'_{\parallel}, {\bf p}'_T), p '_{\parallel}$ and ${\bf
p}'_T$ are projections of its momenta parallel and  perpendicular to beam direction.
 The contribution to the inclusive
cross section of the diagram of Fig.1 is equal
$$ E'\int \frac{d\sigma_{weak}}{d^3 p'} d\eta'\biggl \vert_{\mid \eta'\mid < \eta}
= \frac{9}{8}~ \frac{g^4}{(2\pi)^2}~\frac{p_T}{E}\int ~\frac{dx_1dx_2dx_3}{x_1x_2}
~\frac{(x^2_1 +x^2_2)(1+th^2 \eta)}{(4E^2x_1x_2 -m^2_W)^2 + m^2_W\Gamma^2_W}\times
$$ \be  \frac{1}{(x_1+x_2)x_3 sech~\eta -\frac{p_T}{E}} ~ F_u(x_1) F_d(x_2)\sum_i
\biggl [D^i_u (x_3) +D^i_d (x_3) + D^i_s(x_3) +D^i_c(x_3)\biggr ].\label{4}\ee Here  $E$
is the proton or antiproton energy in c.m.s., $m_W$ and $\Gamma_W$ are $W$ mass and
width, $F_u(x_1),F_d(x_2)$ -- are $u$ and $d$-quark distributions in proton,
$D^i_u,D^i_d, D^i_s, D^i_c$ are the fragmentation functions of $u,d,s,c$ quarks into
$i$-th charged particle, the sum is performed  over all charged particles.  The
integration domain in variables $x_1,x_2,x_3$ is restricted by
$$ x_1x_2 > \biggl ( \frac{p_T}{E}\biggr )^2$$
\be (x_1 +x_2) x_3 sech~\eta > \frac{p_T}{E} (1+x_3 sech~\eta)\label{5}\ee

As well known, (see e.g. \cite{Khoze} and references herein) in case of production of
narrow vector resonances (like $\omega, \varphi, J/\psi, \Upsilon)$ in
$e^+e^-$-annihilation $e^+e^- \to V$ the radiative effects due to emission of real or
virtual photons by initial $e^+$ and $e^-$ are very important. The resonance curves are
widened and resonance maxima are suppressed. The cross section  of the process $e^+e^-
\to V$ without radiative effects is described by Breit-Wigner formula
\be \sigma(e^+e^-\to V) = \frac{12\pi}{s} \frac{\Gamma^V_{e^+e^-}}{M} \mbox{Im}
f_0(\sqrt{s}),\label{6}\ee where $\Gamma^V_{e^+e^-}$ is the electron width of
$V$-resonance, $M$ -- s its mass,
\be f_0(\sqrt{s}) =\frac{(1/2)M}{-\sqrt{s} +M -i\Gamma/2}, \label{7}\ee $\sqrt{s}$ is the
total energy of $e^+e^-$ pair in their c.m.s. and $\Gamma$ is $V$ total width. In
\cite{Khoze} is was shown, that the account of radiative effects results to substitution
\be f_0(\sqrt{s}) \to f(\sqrt{s}) = [~f_0(\sqrt{s})~ ]^{-\beta_{QED}}\label{8}\ee where
\be \beta_{QED} = \frac{4\alpha}{\pi} \biggl [ \ln \frac{\sqrt{s}}{m_e}
-\frac{1}{2}\biggr ],\label{9}\ee $\alpha= e^2 = 1/137,~m_e$ -- is the electron mass.

The similar situation takes place in production of $W$ or $Z$ bosons by annihilation of
quark pair. The diagrams  corresponding  to gluon corrections of the first order are
shown on Fig.2.

\begin{figure}[h]
\hspace{5mm} \epsfxsize=14.0cm \epsfbox{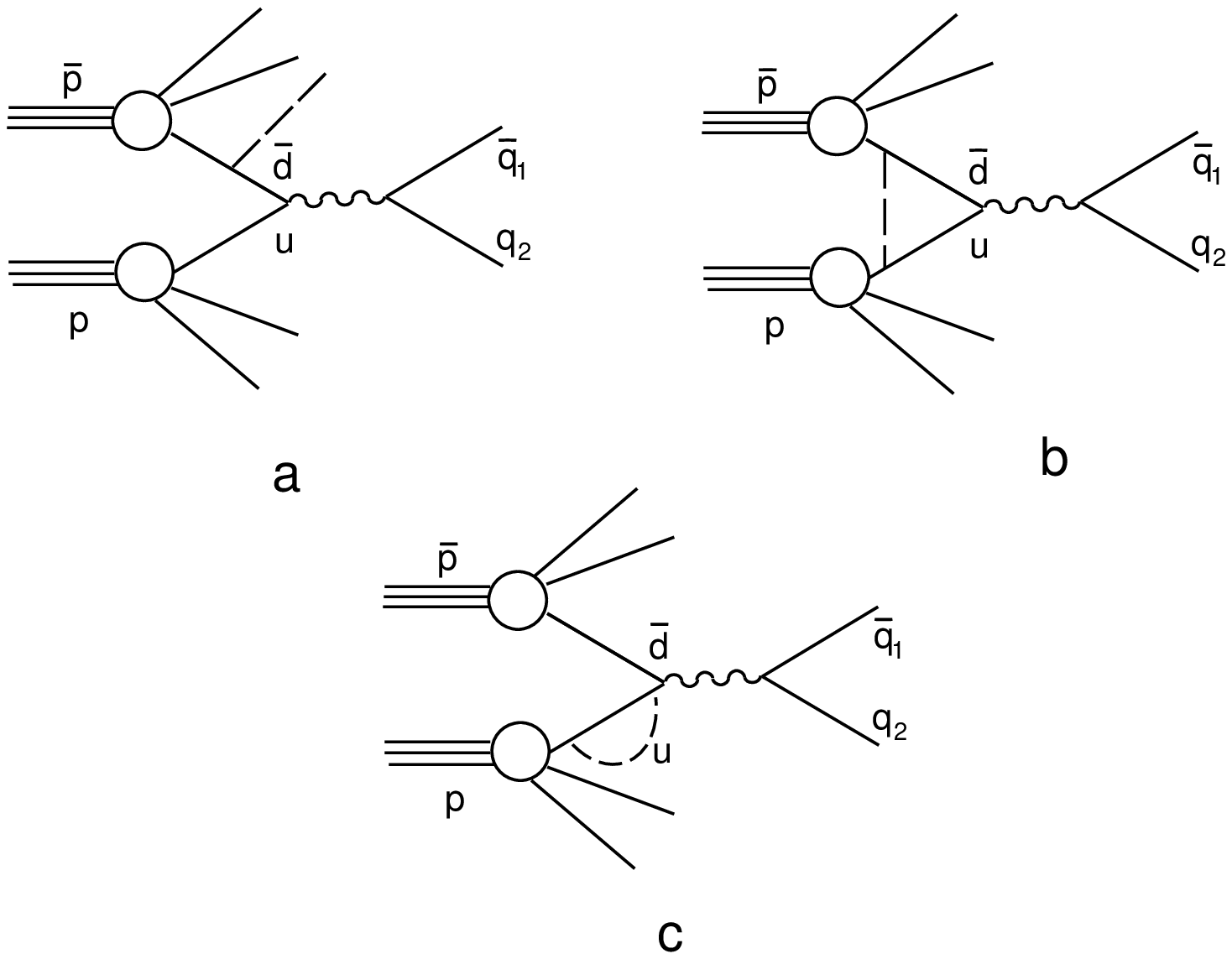}

\vspace{3mm}

 {\bf Fig.2.} Gluon corrections to the process $\bar{q}q \to W$. Dashed  lines correspond
 to gluons.


\end{figure}
The diagrams Fig.2,b,c contribute through interference with the diagram of Fig.1. The
gluon emission by final quarks $\bar{q}_1, q_2$ can be neglected, since these quarks are
currying large $p_T$.
 In analogy with (\ref{8}),(\ref{9}) the account of gluon corrections results
to substitution in Eq.4:
\be \frac{1}{(4x_1x_2 E^2 -m^2_W)^2 + m^2_W\Gamma^2}\to \frac{1}{m^3_W \Gamma_W}~\Biggl [
\frac{m^3_W \Gamma_W}{(4x_1x_2 E^2 - m^2_W)^2 + m^2_W \Gamma^2_W}\Biggr
]^{1-\beta_{QCD}}\label{10}\ee where
\be \beta_{QCD} =\frac{8}{3} \frac{\alpha_s(s)}{\pi} \biggl [\ln \frac{s}{M^2_{char}}
-1\biggr ],~~~s = 4x_1x_2 E^2, \label{11}\ee and $M_{char}$ is the characteristic mass of
strong interaction,  $M_{char} \sim 1$ GeV. At the derivation of (\ref{10}) it was
assumed that $\alpha_s(2p_T)$ is small, $\alpha_s(s) \ll 1$, the $\ln[s/m^2_{char}]$ is
large, the product $\alpha_s(s)\ln[s/M^2_{char}]$ is of order of 1 and the terms $\sim
(\alpha_s\ln[s/M^2_{char}])^n$ are summed. With account of gluon corrections we have
instead of (\ref{4}):
$$ E'\int \frac{d\sigma_{weak}}{d^3p'} d\eta' \biggl \vert_{\mid \eta'\mid < \eta}
=\frac{9}{8} \frac{g^2}{(2\pi)^2} \frac{p_T}{E} \int \frac{dx,dx_2,dx_3}{x_1x_2}
\frac{(x^2_1+x^2_2)(1+th^2\eta)}{m^3_W\Gamma_W}\times$$
$$\times \biggl [ \frac{m^3_W
\Gamma_W}{(4x_1x_2 E^2- m^2_W)^2 + m^2_W \Gamma^2_W}\biggr ]^{1-\beta_{QCD}}
\frac{1}{(x_1+x_2)x_3 sech~\eta - p_T/E} F_u(x_1) F_d(x_2) \times$$

\be \times \sum_1 [D^i_u(x_3) + D^i_d(x_3) + D^i_s(x_3) + D^i_c(x_3)]\label{12}\ee At
$p_T\approx 40-100$ GeV $\alpha_s(2p_T)= 0.12 -0.10$ and $\beta_{QCD} \approx 0.6-0.8$.
So, the account of gluon corrections drastically  changes the results. At small $x$ the
quark distributions and fragmentation functions behave as $(1/x)^{\gamma}$, where
$\gamma$ is equal or larger  than 1 and small $x_1,x_3$ or  $x_2,x_3$ are dominating in
(\ref{12}). The consequence of this fact is that the cross section (\ref{12}) increases
with beam energy $E$ and decreases with $p_T$ more slowly, than $1/p^4_T$.
 Therefore the measurements of inclusive cross sections at high $p_T$
at LHC are very promisable. The detailed calculation of weak contribution to the
inclusive cross section at high $p_T$, using quark distributions found by MSTW2008
\cite{Martin} and CTEQ 6.6.M \cite{CTEQ} and available information on distribution
functions will be presented in separate publication.

\begin{figure}[h]
\hspace{30mm} \epsfxsize=11.0cm \epsfbox{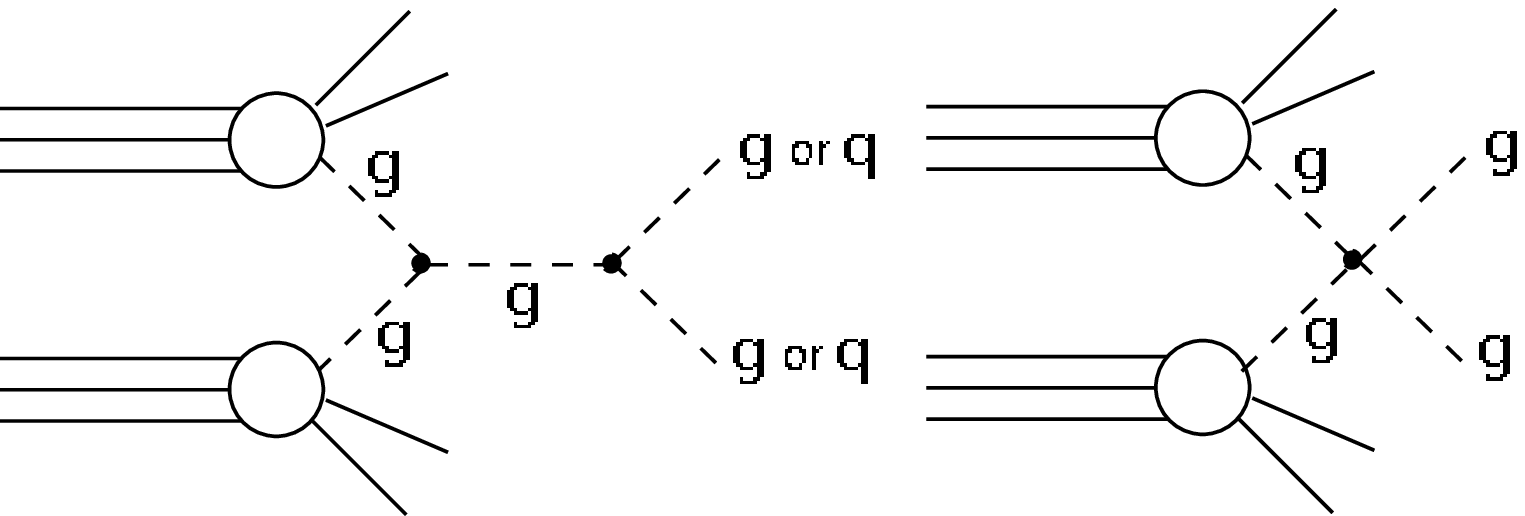}


\vspace{3mm}

{\bf Fig.3.} The diagrams representing the contributions to inclusive jets production due
to gluon exchange.
\end{figure}

 Finally, let us now explain,  why experimentally scaling
violation is not observed in inclusive jet production \cite{Aaltonen},\cite{Abazov}. In
this case in strong interaction mechanism the main role are playing the diagrams with
gluon exchange, like ones presented on Fig.3 and the contribution of weak interaction is
small in comparison with them.


I am thankful to A.B.Kaidalov for useful remark and to A.G.Oganesian who  found few
errors in my calculations.

This work was supported in parts by RFBR grant 09-02-00732 and by CRDF Cooperative
Program grant RUP2-2961-MO-09. The support of the European Community -- Research
Infrastructure  Integrating  Activity ``Study of Strongly Interacting Matter'' under
Seventh Framework Program of EU is acknowledged.


\vspace{5mm}


\begin{thebibliography}{99}
\bibitem{CDF}  CDF Collaboration, T.Aaltonen et al, Phys.Rev {\bf 79}, 112005 (2009)
\bibitem{Ioffe} B.L.Ioffe, Phys.Lett. {\bf 30B} 123 (1969).
\bibitem{Albino} S.Albino, B.A.Kniehl and G.Kramer, arXiv: Phys.Rev.Lett.
{\bf 104}, 242001 (2010), hep-ph/1003.2127
\bibitem{Gribov} V.N.Gribov, B.L.Ioffe and I.Ya.Pomeranchuk, Yad.Phiz. {\bf 2}, 768
(1965), Sov.J.Nucl.Phys. {\bf 2}, 549 (1966).
\bibitem{Yoon} A.S.Yoon, E.Wenger and G.Roland, arXiv: hep-ph/1003.5928
\bibitem{Aaltonen} CDF Collaboration, T.Aaltonen et al, Phys.Rev {\bf 78}, 052006 (2008),
ibid {\bf 79}, 119902(E) (2009) [arXiv: hep-ph/0807.2204]
\bibitem{Abazov} D0 Collaboration, V.M.Abazov et al, Phys.Rev.Lett. {\bf 101}, 06001
(2008)

[arXiv: hep-ph/0802.2400]
\bibitem{Khoze} B.L.Ioffe, V.A.Khoze and L.N.Lipatov, Hard Processes, North Holland,
1984, Chapter 2, Sec. 2.11.2.
\bibitem{Martin} A.D.Martin, W.J.Stirling, R.S. Thorne and G.Watt, Eur.J.C. {\bf 63}, 189
(2009)

[arXiv: hep-ph/0901.0002]
\bibitem{CTEQ} CTEQ Collaboration, P.M.Nadolsky et al, Phys.Rev {\bf D78} 013004 (2008)

[arXiv: hep-ph/0802.0007]



\end{thebibliography}
\end{document}